\documentclass[twocolumn,showpacs,preprintnumbers,
amsmath,amssymb]{revtex4}
\usepackage{graphicx}

\begin{document}
\title{Bose-Einstein Condensation Picture of Superconductivity in
$\textrm{Ag}_{2}(\textrm{Ag}_{3}\textrm{Pb}_{2}\textrm{H}_{2}\textrm{O}_{6}),$
$\textrm{Na}_{0.05}\textrm{WO}_{3}$ and
$\textrm{Na}_{0.041}\textrm{NH}_{3}$ composites. (Dilute metals).}
\author{V. N. Bogomolov}
\affiliation{A. F. Ioffe Physical \& Technical Institute,\\
Russian Academy of Science,\\
194021 St. Petersburg, Russia}
\email{V.Bogomolov@mail.ioffe.ru}
\date{\today}
\begin{abstract}
Traditionally, when one describes the crystallographic structure of oxides, the oxygen ion radius
 $r_{02-}$ is assumed to be approximately equal to $1.4\AA.$   The oxygen ions occupy in this case
 $80-90\%$
 of the crystal volume. Metal atoms are considered then as ions playing a role of donors with rather
 small radius of $(0.5 -- 0.8)\AA.$ However, the atomic packing picture and, therefore, physical properties
 such as electric conductivity and superconductivity of oxides will be essentially different, if we
 assume $r_{02-}\sim0.56 \AA.$ Such magnitude of the radius is known from the quantum mechanics calculations \cite{bib2}.
 According to this picture, $80-90\%$ of the crystal volume is occupied by the metal atom orbitals
 with radius $(1.3-1.9)\AA,$  while the oxygen ions play a role of acceptors, which reduce occupancy
 of these orbitals ("indirect" dilution of the metal).    A "direct" dilution of metals takes place
 in stoichiometric matrices. When $r_{02-}~0.56 \AA,$ channels with diameter $3.6$   present in the hexagonal
 matrix $\textrm{Ag}_{3}\textrm{Pb}_{2}\textrm{H}_{2}\textrm{O}_{6}$ directed along the "\textit{c}" axis.
 The channels are filled by chains of the $\textrm{Ag}_{2}$ molecules
 with atomic diameter of $2.6\AA$ and the molecule concentration  $n_{B} = 25.6\times10^{20} cm^{-3}.$
 The Bose-Einstein condensation (BEC) temperature $T_{cB}\sim400\textrm{K}$ is calculated
 for the electron effective mass value $m^{*}=7.5m_{e},$
 where $m_{e}$ is the isolated electron mass. Three-dimensional networks of $\textrm{Na}_{2}$ chains form
 in Na solutions
 in $\textrm{NH}_{3}$ and $\textrm{WO}_{3}$ as well (respectively, with $n_{B}=4.71\times10^{20}cm^{-3};$
 $T_{c}\sim180\textrm{K};$ $ m^{*}=5.0m_{e}$ and $n_{B}=4.74 \times10^{20}cm^{-3};$
$T_{c}\sim91\textrm{K};$ $ m^{*}=10m_{e}).$ Close magnitudes of
the $\textrm{Ag}_{2}$ and $\textrm{Na}_{2}$ chains parameters
respectively in $\textrm{NH}_{3}$ and $\textrm{WO}_{3}$  favors
the opinion that all these structures have a composite structure
and similar mechanisms of the high temperature superconductivity.
\end{abstract}
\pacs{71.30.+h, 74.20.-z, 74.25.Jb}
\maketitle
\bigskip
   A conductivity of $\textrm{Ag}_{2}(\textrm{Ag}_{3}\textrm{Pb}_{2}\textrm{H}_{2}\textrm{O}_{6}),$
   significantly higher than conductivity of
   pure Ag in the temperature range of $300-500\textrm{K}$ was observed in \cite{bib1}.
   As high as this the conductivity magnitude is possible only if at least a part of the sample volume
   is in the superconducting state. However, the quasi one-dimensional nature of this crystal makes
   confirmation of the superconducting state by means of magnetic measurements rather difficult.
   If one will use, instead of the traditional value of $1.4\AA,$ the  value of $0.56\AA$
   for the \!$\textrm{O}^{2-}$\!\!
   ion radius, which was obtained in result of quantum mechanical calculations \cite{bib2}, the concept
   the crystal structure, packing of atoms, conductivity and superconductivity of the oxide will
   change essentially and they will appear to be similar to properties of the nano-composites
   $\textrm{Na}_{0.05}\textrm{WO}_{3}$ \cite{bib3} and $\textrm{Na}_{0.041}\textrm{NH}_{3}$ \cite{bib4}.

    Packing of atoms in the   $\textrm{Ag}_{2}(\textrm{Ag}_{3}\textrm{Pb}_{2}\textrm{H}_{2}\textrm{O}_{6})$
compound according to the X-ray analysis data
\cite{bib1,bib5,bib6} is depicted in Fig.\textit{a}.
 Pb atoms have an
intermediate valence, which equals $3.5$ in the present case
$(\textrm{Pb}^{3+}--\textrm{Pb}^{4+}).$ The following values of
radii were used: $r_{02-}\sim0.56 \AA.$   and $r_{\textrm{Ag}} =
1.30\AA,$ $ r_\textrm{{Pb}} = 1.65\AA,$ which are near to the
related metal atomic radius values. The $\textrm{Ag}_{2}$ molecule
concentration $n_{B} = 25.6\times10^{20} cm^{-3}.$ A view along
the \textit{c}-axis onto one atomic layer is presented in
Fig.\textit{a}. Darkened circles stand for Ag atoms forming
molecular chains with the interatomic distance of $3\AA$
\cite{bib6}, which fill the channels directed along the
\textit{c}-axis with diameter $3.6 \AA.$ Having the diameter
$2.6\AA,$  the molecular chains are well isolated from the
$\textrm{Ag}_{3}\textrm{Pb}_{2}\textrm{H}_{2}\textrm{O}_{6}$
matrix. Interphase space of such scale $(h\sim0.5\AA -1.0\AA)$ was
experimentally observed in the metal-zeolite systems \cite{bib7}.
A view onto the (\textit{a-c)} face of the unit cell is shown in
Fig.b. A lattice of the $\textrm{Ag}_{2}$ quasi one-dimensional
chains forms an anisotropic $3D$ medium ("gossamer" type, see
Fig.c.)

    BEC occurs in the ideal Bose gas, when the de Broglie wavelength satisfies the relation

  $\lambda_{B}=(2.612/n_{B})^{1/3}=h/(2\pi MkT_{cB})^{1/2},$     or

  $T_{cB} = 3.31 (h/2\pi)^{2}\, n_{B}^{2/3}/M.$

Assuming that BEC occurs in the array of the $\textrm{Ag}_{2}$
chains at $T_{cB}\sim400\textrm{K},$ we can estimate the electron
effective mass in chains $M/2 = m^{*}:$

 1)  For $n_{B}=24.6\times10^{20} cm^{-3}$ and
estimated $T_{c}\sim400\textrm{K}$ \cite{bib1}, we obtain
$M=2m^{*}=15m_{e},$ or $m^{*}=7.5 m_{e}.$

 2) It is seen from Fig.c.
that overlapping of the electron pairs of the $\textrm{Ag}_{2}$
molecules is stronger along the chains, than between ones.
However, $T_{c}$ is determined by the distance between the
molecule centers (X), which equals $\sim 6\AA
\sim\lambda_{B}/1.38\AA \sim 7.4\AA.$

3) Anisotropy of the molecular pairs seemingly shows itself in
anisotropy of the superconducting state of the quasi
one-dimensional superconductor
$\textrm{Ag}_{2}(\textrm{Ag}_{3}\textrm{Pb}_{2}\textrm{H}_{2}\textrm{O}_{6})$.

Arguments in favor of forming of the\,$3D$\,network of
$\textrm{Na}_{2}$ molecule chains in the insulating matrix of
$\textrm{Na}_{0.05}\textrm{WO}_{3}$ \cite{bib3} and
$\textrm{Na}_{0.041}\textrm{NH}_{3}$ \cite{bib4} (irregular
"cobweb") are presented below.

 Concentration of the $\textrm{Na}_{2}$ molecules is also known. It equals  $4.71\times10^{20}cm^{-3}$
 in $\textrm{Na}_{0.041}\textrm{NH}_{3} (T_{cB}\sim 180\textrm{K})$ \cite{bib4}, and $4.74\times10^{20}cm^{-3}$
 in $\textrm{Na}_{0.05}\textrm{WO}_{3} (T_{cB}~ 91\textrm{K})$ \cite{bib3}. Assuming that BEC takes place,
 we conclude that the electron effective mass magnitudes in  $\textrm{Na}_{2}$ chains in these
 nano-composites (respectively, $5m_{e}$ and $10m_{e}$) are near to $m^{*}$ value in $\textrm{Ag}_{2}$
 chains $(7.5 m_{e}).$
 Distinction of the mass values can be a result of the dielectric constant
 difference in the insulators $\textrm{NH}_{3}$ and $\textrm{WO}_{3}$
 (respectively, $16$ and $35$ at $300\textrm{K})$ and $\textrm{Ag}_{3}\textrm{Pb}_{2}\textrm{H}_{2}\textrm{O}_{6}.$

 Not only closeness of the effective mass values in all of three materials, but a nature of the binding
  energy $\Delta$ of these pairs can be considered as an evidence in favor of the chain structure for
 $\textrm{Na}_{2}$ molecules in $\textrm{Na}_{0.041}\textrm{NH}_{3}$ and $\textrm{Na}_{0.05}\textrm{WO}_{3}$
 as well. The molecular chains can be one of the steps of the phase separation process in the composites
 (atoms - molecules - chains - clusters...).

   The BEC transition temperature $T_{c}$ for the molecular chain structure can be estimated
   from the energetic consideration. If the atomic chain with the binding energy \textit{Q} is stretched
 (by the matrix, for instance) up to the interatomic distance $D_{1},$ and subsequently atoms comprising
 the chain pair off forming diatomic molecules with the interatomic distance $D_{2},$ the binding energy
 of the diatomic molecules reads
         $$ \Delta\sim (D_{1}-D_{2})Q/D_{2}  \sim T_{cB}.$$
  This relation is very good satisfied for a variety of  fullerides \cite{bib8}.

1)  In the case of
$\textrm{Ag}_{2}(\textrm{Ag}_{3}\textrm{Pb}_{2}\textrm{H}_{2}\textrm{O}_{6})$
all parameters are known: $Q=2.8\,\textrm{eV}; \quad
D_{1}=3.2\,\AA;\quad D_{2}=3.0\,\AA;$ therefore,
$\Delta=0.187\textrm\,{eV}.$ From the relation  ${\Delta \sim 5.5
kT_{cB}}$ \cite{bib9} it follows  that $T_{c} =395\,\textrm{K},$
which is very near to the estimates in \cite{bib1}.

2)  The unit cell of $\textrm{Na}_{0.05}\textrm{WO}_{3}$ has a
orthorhombic symmetry ${(a = 3.67\times2 \AA; \quad b = 3.73 \AA;
\quad c = 3.85 \AA)}$ \cite{bib10}. Since the Na atoms diameter
${d_{2}=3.71\AA < 3.85 \AA   = D_{1},}$ the $\textrm{Na}_{2}$
molecules are directed along the c-axis. We have $\Delta=0.043
\textrm{eV}$ and $T_{c} = 88.8 \textrm{K}$ for $Q=1.17
\textrm{eV}$ that is almost in conformity with experimental
results $(91\textrm{K})$ \cite{bib3}.

3) The Na solution in $\textrm{NH}_{3}$ is metastable and
superconductivity with $T_{c} \sim 180\textrm{K}$ was observed
(using the Meissner effect) only a few times from $200$ attempts
 \cite{bib4}. Forming of $\textrm{Na}_{2}$ molecules means beginning of the
phase separation. The Na atomic volume in $\textrm{NH}_{3}$ is
known to be larger, than in the metal \cite{bib4}. The interatomic
distance increase in molecules is related to a decrease of the
binding energy for the metal bonding. The binding energy of the
isolated $\textrm{Sn}_{2}$ molecules equals $\sim 0.35Q$
\cite{bib11,bib12,bib13}. Therefore, $T_{c} \sim 180 \textrm{K}$
for $d_{2} = 3.71\AA$ could be related to the distance $D_{2} \sim
4\AA.$

  The most reliable calculations can be done, to be sure only for
  $\textrm{Ag}_{2}(\textrm{Ag}_{3}\textrm{Pb}_{2}\textrm{H}_{2}\textrm{O}_{6})$.
  They are only qualitative in other cases, but the scale of magnitudes seems to reasonable.
   A quantitative relation between $\Delta$   and $T_{cB}$ can be found from the simple consideration as well.
   Tunneling of the molecular electronic pairs can lead to their BEC. But increase of $n_{B}$ leads to the drop
   of the electron pair binding energy $\Delta$  and at some value of $n_{cB}$ it is energetically favorable
   for electrons to pass into the Fermi state (metal) or into the insulator state. Equating the electron
   energy density in a metal and in the Bose condensate we have:
      $$E_{F^{*}}\sim2.87(h/2\pi)^{2}n_{cB}^{5/3}/m^{*} \sim \Delta
      n_{cB}/2,$$
That gives the estimate for the BEC - BCS  crossover:$\Delta \sim
5.5 kT_{cB}$ \cite{bib9}.  Besides,\\
$\lambda_{B}=(2.612/n_{B})^{1/3} = h/(2 \pi MkT_{cB})^{1/2} =
h/(2m^{*}6.28 kT_{cB})^{1/2} \sim h/(2m^{*}\Delta)^{1/2}; \quad
m^{*}=M/2$
  or     $\lambda_{B}=(2.612/n_{B})^{1/3} \sim h/(2m^{*}\Delta)^{1/2}$    (the single electron tunneling condition).

Conclusions:

   Similarity of parameters of $\textrm{Ag}_{2}$ and $\textrm{Na}_{2}$ chains in essentially
   distinct matrices $\textrm{Ag}_{3}\textrm{Pb}_{2}\textrm{H}_{2}\textrm{O}_{6},\qquad \textrm{WO}_{3}$
   and $\textrm{NH}_{3}$ indicates the chain composite structure of these materials and likeness of
   the phase transition mechanisms (BEC mechanism of the high temperature superconductivity).
   There exists a $3D$ network of the $\textrm{Me}_{2}$ chains in the form of "gossamer" or  "cobweb"
   structures with circular closed loops \cite{bib14,bib15} demonstrating superconductivity of the BEC type.

   We can formulate the material requirements for development of promising for practice new superconductors.

1)  Essential three-dimensionality of high temperature
superconductors is necessary.

2)  High magnitude of $\Delta$  is good (for instance, Na can be
substituted by Li, Ag or divalent atoms).

3) Maximum concentration of $\textrm{Me}_{2}$ molecules or of
other particles with electron pairs is desirable.

   Approaching of such magnitudes of the parameters can be a major physical-chemical and methodical challenge.
The $\textrm{Na}_{0.25}\textrm{WO}_{3}$ system, for example, is
located at the margin of the structure transition into the
stoichiometric compound (tungsten bronze)
$\textrm{Me}-\textrm{WO}_{3n-1}$ with the electrons in the Fermi
state. There are a lot of matrices having nano-size voids
(zeolites, for instance), which can be used for developing of such
nano-composites. Chains or "gas" of particles with electron pairs
can be stabilized by the matrix near the threshold of the
insulator - superconductor transition.

 Similar problems can be encountered in consideration of stoichiometric compounds
 like YBaCuO. But in this case the situation is not so clear since the effective
 density of electron pairs is determined by occupation of the metal atoms orbitals: "diluted" metals \cite{bib9}.

\begin{figure}[tbp]
\end{figure}
\includegraphics[width=\linewidth]{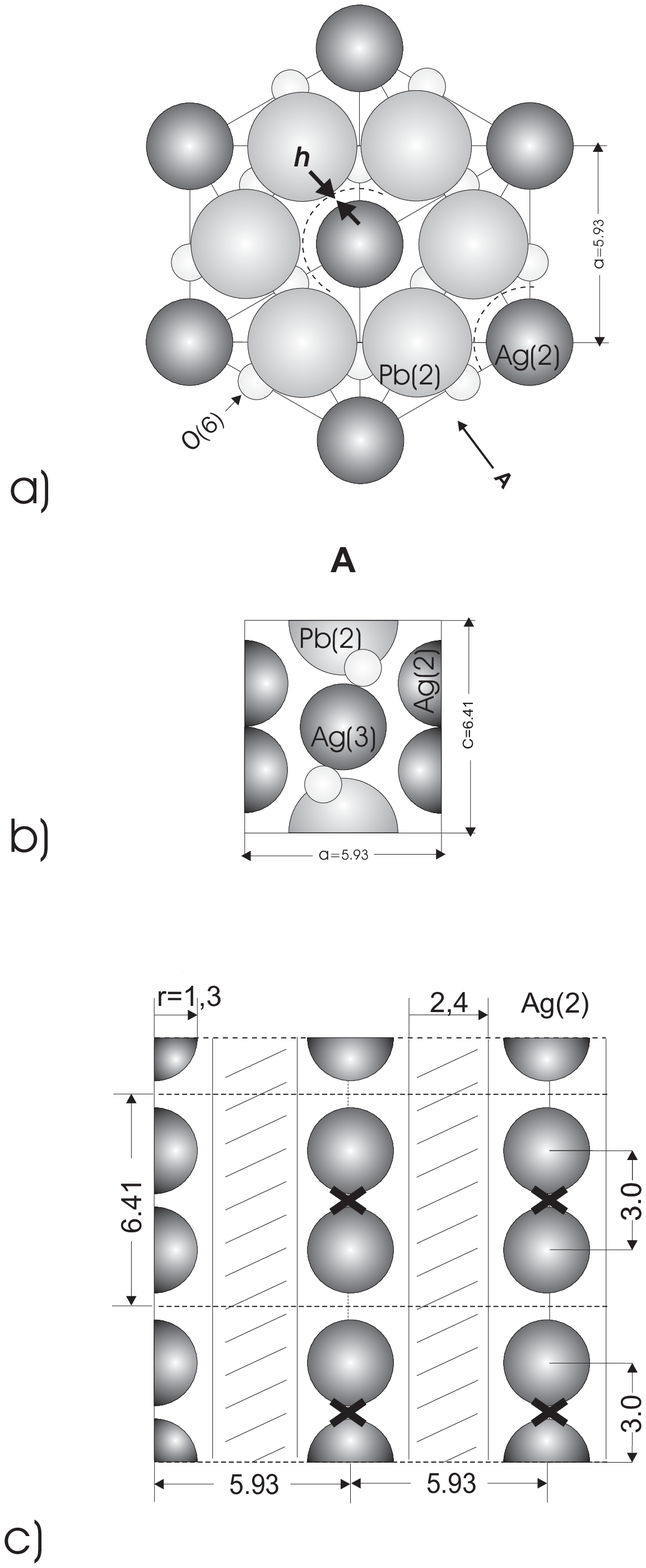}

\end{document}